\documentstyle[12pt]{article}
\begin{document}
\begin{titlepage}
\begin{center}
{\Large \bf  Exploring Softly Broken SUSY Theories
\protect\\[0.5cm] via Grassmannian Taylor Expansion}

\vglue 10mm {\bf  D.I.~Kazakov }

\vglue 5mm {\it Bogoliubov Laboratory of Theoretical Physics, \\
Joint Institute for Nuclear Research, \\ 141 980 Dubna, Moscow
Region, RUSSIA\\ e-mail: kazakovd@thsun1.jinr.ru}
\end{center}

\vglue 20mm
\begin{abstract}
We demonstrate that soft SUSY breaking introduced via replacement
of the couplings of a rigid theory by spurion superfields has far
reaching consequences. Substituting these modified couplings into
renormalization constants, RG equations, solutions to these
equations, fixed points,  finiteness conditions, etc., one can get
corresponding relations for the soft terms by a simple Taylor
expansion over the Grassmannian variables. This way one can get
new solutions of  the RG equations. Some examples including the
MSSM, SUSY GUTs and  the N=2 Seiberg-Witten model are given.
\end{abstract}

\vglue 10mm \noindent PACS numbers: 11.10Gh, 11.10Hi, 11.30Pb \\
Keywords: soft supersymmetry breaking, renormalization,
renormalization group

\end{titlepage}

\section{Introduction}

In a recent paper~\cite{AKK}, which is based  on the previous
publications~\cite{spurion, Yamada} we have shown that
renormalizations in a softly broken SUSY theory follow  from those
of an unbroken SUSY  theory  and can be performed in the following
straightforward  way:

{\it One takes  renormalization constants of a rigid theory,
calculated in some massless scheme, substitutes instead of the
rigid couplings (gauge and Yukawa) their modified expressions,
which depend on a Grassmannian variable, and expand over this
variable.}

This gives   renormalization constants for the soft terms.
Differentiating them with respect to a scale one can find
corresponding renormalization group equations.

Thus the soft term renormalizations are not independent but can be
calculated from the known renormalizations of a rigid theory with
the help of the differential operators. Explicit form of these
operators has been found in a general case and in some particular
models like SUSY GUTs or the MSSM~\cite{AKK}. The same expressions
were obtained also in ref.~\cite{NewJ}.

In this letter we demonstrate that this procedure works at all
stages. One can make the above mentioned substitution on the level
of the renormalization constants, RG equations, solutions to these
equations, approximate solutions, fixed points, finiteness
conditions, etc. Expanding then over a Grassmannian variable one
obtains  corresponding expressions for the soft terms. This way
one can get new solutions of  the RG equations and explore their
asymptotics, or approximate solutions, or find their stability
properties, starting from the known expressions for a rigid
theory.

Below we give some examples and in particular consider the MSSM
with low $\tan\beta$, where  analytical solutions are known. We
show how one can easily obtain solutions to the RG equations for
the soft mass terms much simplier than are known in the
literature.  In a finite SUSY GUT finiteness conditions for the
soft terms  appear as a trivial consequence of a finiteness of a
rigid theory. Another example is the N=2 SUSY model, where the
exact (non-perturbative) Seiberg-Witten solution is known. Here
one can extend the S-W solution to the soft terms.

\section{Soft SUSY Breaking  and Renormalization}

Consider an arbitrary $N=1$ SUSY gauge theory with unbroken SUSY.
The Lagrangian of a rigid theory is given by
\begin{eqnarray}
{\cal L}_{rigid} &=& \int d^2\theta~\frac{1}{4g^2}{\rm
Tr}W^{\alpha}W_{\alpha} + \int d^2\bar{\theta}~\frac{1}{4g^2}{\rm
Tr} \bar{W}^{\alpha}\bar{W}_{\alpha}.  \label{rigidlag} \\ &+&
\int d^2\theta  d^2\bar{\theta} ~~\bar{\Phi}^i (e^{V})^j_i\Phi_j +
\int
 d^2\theta ~~{\cal W} + \int d^2\bar{\theta} ~~\bar{\cal W},  \nonumber
\end{eqnarray}
where $W^{\alpha}$ is the field strength chiral superfield and the
superpotential ${\cal W}$  has the form
\begin{equation}
{\cal  W}=\frac{1}{6}\lambda^{ijk}\Phi_i\Phi_j\Phi_k +\frac{1}{2}
M^{ij}\Phi_i\Phi_j.\label{rigid}
\end{equation}

To  perform the SUSY breaking, which satisfies the requirement of
"softness", one can introduce a gaugino mass term as well as cubic
and quadratic interactions of  scalar superpartners of the matter
fields~\cite{spurion}
\begin{eqnarray}
-{\cal L}_{soft-breaking} &=&\left[ \frac{M}{2}\lambda\lambda
+\frac 16 A^{ijk} \phi_i\phi_j\phi_k+ \frac 12 B^{ij}\phi_i\phi_j
+h.c.\right] +(m^2)^i_j\phi^{*}_i\phi^j,\label{sofl}
\end{eqnarray}
where $\lambda$ is  the gaugino field and $\phi_i$ is the lower
component of the chiral matter superfield.

One can rewrite  the Lagrangian (\ref{sofl}) in terms of  N=1
superfields introducing  the external spurion
superfields~\cite{spurion} $\eta=\theta^2$ and $\bar
\eta=\bar{\theta}^2$, where $\theta$ and $\bar \theta$ are
Grassmannian parameters, as~\cite{Yamada}
 \begin{eqnarray} {\cal
L}_{soft} &=& \int d^2\theta~\frac{1}{4g^2}(1-2M\theta^2) {\rm
Tr}W^{\alpha}W_{\alpha} + \int
 d^2\bar{\theta}~\frac{1}{4g^2}(1-2\bar{M}\bar{\theta}^2) {\rm
Tr}\bar{W}^{\alpha}\bar{W}_{\alpha}.  \nonumber \\ &+&\int
d^2\theta d^2\bar{\theta} ~~\bar{\Phi}^i(\delta^k_i -(m^2)^k_i\eta
\bar{\eta})(e^V)^j_k\Phi_j   \label{ssofl2} \\ &+& \int  d^2\theta
\left[\frac 16 (\lambda^{ijk}-A^{ijk} \eta)\Phi_i\Phi_j\Phi_k+
\frac 12 (M^{ij}-B^{ij}\eta ) \Phi_i\Phi_j \right] +h.c.
\nonumber
\end{eqnarray}

Comparing eqs.(\ref{rigidlag}) and (\ref{ssofl2}) one can see that
eq.(\ref{ssofl2}) is equivalent to (\ref{rigidlag}) with
modification of the rigid couplings $g^2, \lambda^{ijk}$ and
$M^{ij}$, so that they become external superfields dependent on
Grassmannian parameters $\theta^2$ and $\bar{\theta}^2$. The
scalar mass term $m^2\eta\bar \eta$ modifies  fields $\Phi$ and
$\bar{\Phi}$. These modifications of the couplings and fields are
valid not only for the classical Lagrangian but also for the
quantum one.\footnote{Throughout the paper   the existence of some
SUSY invariant regularization is assumed.} As has been shown in
Ref.~\cite{AKK} the following statement is valid:

{\it If a rigid theory {\rm (\ref{rigidlag}, \ref{rigid})} is
renormalized via introduction of  renormalization constants $Z_i$,
defined within some minimal subtraction massless scheme, then a
softly broken theory {\rm (\ref{ssofl2})} is renormalized via
introduction of  renormalization superfields $\tilde{Z}_i$ which
are related to $Z_i$ by the coupling constants redefinition
\begin{equation}
\tilde{Z}_i(g^2,\lambda ,\bar \lambda)
 =Z_i(\tilde{g}^2,\tilde{\lambda},\tilde{\bar \lambda}),
\label{Z}
\end{equation}
where the redefined couplings are
\begin{eqnarray}
\tilde{g}^2&=&g^2(1+M \eta+\bar M \bar{\eta}+2M\bar M \eta
\bar{\eta}),\ \ \  \eta=\theta^2, \ \ \ \bar{\eta}=\bar{\theta}^2,
\label{g}\\ \tilde{\lambda}^{ijk}&=&\lambda^{ijk}-A^{ijk}\eta
+\frac 12 (\lambda^{njk}(m^2)^i_n
+\lambda^{ink}(m^2)^j_n+\lambda^{ijn}(m^2)^k_n)\eta \bar \eta,
\label{y1}\\ \tilde{\bar \lambda}_{ijk}&=&\bar \lambda_{ijk} -
\bar A_{ijk} \bar{\eta}+ \frac 12 (\bar \lambda_{njk}(m^2)_i^n
+\bar \lambda_{ink}(m^2)_j^n+\bar \lambda_{ijn}(m^2)_k^n)\eta \bar
\eta
 .  \label{y2}
\end{eqnarray} }

Thus a softly broken SUSY gauge theory is equivalent to an
unbroken one in external spurion superfield as far as the
renormalization properties are concerned. Substitutions
(\ref{g}-\ref{y2}) can be made not only in the renormalization
constants, but at every stage of  the renormalization procedure,
since the RG functions and RG equations are derived from
renormalization constants applying the differential operators. The
key point is that one can consider an unbroken  theory in external
superfield which is equivalent to replacing of the couplings by
external superfields according to eqs.(\ref{g}-\ref{y2}). Then one
can expand over  Grassmannian  parameters.

In what follows  we would like to simplify the notations and
consider numerical rather than tensorial couplings.   When  group
structure and  field content of the model are fixed, one has a set
of  gauge  $\{g_i\}$ and   Yukawa $\{y_k\}$ couplings.  It is
useful to consider the following rigid parameters
 $$ \alpha_i \equiv \frac{g_i^2}{16\pi^2}, \ \ \  Y_k \equiv
\frac{y_k^2}{16\pi^2}.$$ Then eqs.(\ref{g}-\ref{y2})  look like
\begin{eqnarray}
\tilde{\alpha}_i&=&\alpha_i(1+M_i \eta+\bar M_i
\bar{\eta}+2M_i\bar M_i
 \eta \bar{\eta}), \label{ga}\\
\tilde{Y}_k&=&Y_k(1+A_k \eta +\bar A_k \bar{\eta}+ (A_k\bar
A_k+\Sigma_k) \eta \bar \eta),
 \label{ya}
\end{eqnarray}
where to standardize the notations we have  redefined parameter A:
$ A \to Ay$  in a usual way  and have changed the sign of A to
match it with the gauge soft terms. Here $\Sigma_k$ stands for a
sum of $m^2$ soft terms, one for each leg in the Yukawa vertex.

Now the RG equation for a rigid theory can be written in a
universal form
\begin{equation}
\dot  a_i= a_i\gamma_i(a), \ \ \ \   a_i = \{\alpha_i, Y_k\},
\label{RG}
\end{equation}
where $\gamma_i(a)$ stands for a sum of  corresponding anomalous
dimensions. In the same notation the soft terms
(\ref{ga},\ref{ya}) take the form
\begin{equation}
\tilde{a}_i=a_i(1+m_i \eta+\bar m_i \bar{\eta}+S_i\eta
\bar{\eta}), \label{usoft}
\end{equation}
where  $m_i=\{M_i, A_k\}$ and  $S_i=\{2M_i\bar M_i , A_k\bar
A_k+\Sigma_k \}$.

\section{Grassmannian Taylor Expansion}

We demonstrate now how the RG equations for the soft terms appear
via Grassmannian Taylor expansion from those for the rigid
couplings (\ref{RG}). Indeed, let us substitute  eq.(\ref{usoft})
into eq.(\ref{RG}) and expand over $\eta $ and $\bar \eta$. One
has to be careful, however, since as it follows from the soft
Lagrangian (\ref{ssofl2})  gauge couplings are  involved in chiral
Grassmann integrals and expansion over $\eta$ or $\bar \eta$ makes
sense up to F-terms only.  On the contrary, the Yukawa couplings
$Y$, being a product of $y$ and $\bar y$, are general superfields,
so the expansion is valid for D-terms as well. Having this in mind
one gets
\begin{equation}
\dot{\tilde{a}}_i= \tilde{a}_i\gamma_i(\tilde{a}), \label{RGs}
\end{equation}
Consider first the F-terms.  Expanding over $\eta$  one has
\begin{equation}
\dot  a_i m_i + a_i \dot m_i = a_im_i\gamma_i(a)+
a_i\gamma_i(\tilde{a})\vert_F,
\end{equation}
or
\begin{equation}
 \dot m_i =  \left.\gamma_i(\tilde{a})\right|_F = \sum_j a_j\frac{\partial
  \gamma_i}{\partial a_j}m_j.
\end{equation}
This is just the  RG equation for the soft terms $M_i$ and $A_k$
which was written in Ref.~\cite{AKK} in the form
\begin{equation}
 \dot m_i =  D_1\gamma_i(a).
\end{equation}
Proceeding the same way for the D-terms  one gets after some
algebra
\begin{equation}
 \dot S_i =  \gamma_i(\tilde{a})\vert_D = 2m_i\sum_j a_j\frac{\partial
  \gamma_i}{\partial a_j}m_j+\sum_ja_j\frac{\partial \gamma_i}{\partial a_j}S_j+
   \sum_{j,k} a_ja_k\frac{\partial^2 \gamma_i}{\partial a_j  \partial a_k}m_jm_k.
\label{s}
\end{equation}
Substituting $S_i= m_i\bar m_i+\Sigma_i$ one has the RG equation
for the mass terms
\begin{equation}
 \dot{\Sigma}_i =  \sum_ja_j\frac{\partial \gamma_i}{\partial a_j}(m_jm_j+\Sigma_j)+
  \sum_{j,k} a_ja_k\frac{\partial^2 \gamma_i}{\partial a_j  \partial a_k}m_jm_k.
\end{equation}

One can also obtain  the RG equation for the individual soft
masses out of  field renormalization.  Consider the chiral Green
function in a rigid theory. It obeys  the following  RG relation
\begin{equation}
 <\Phi_i \bar{\Phi}_i> \ \ = \ \ <\Phi_i \bar{\Phi}_i>_0 e^{\displaystyle
  \int_0^t \gamma_i(a(t'))dt'}.
\end{equation}
Making the substitution $$<\Phi_i \bar{\Phi}_i> \ \  \to \ \
<\Phi_i \bar{\Phi}_i>(1+m^2_i\eta\bar \eta),
 \ \ \  a\ \to \ \tilde{a},$$
and expanding over $\eta\bar \eta$ ( since it stands under the
full Grassmann
 integral only  D-term contributes) one has
\begin{equation}
  m^2_i=m^2_{i0} + \int_0^t dt' \left. \gamma_i(\tilde{a}(t'))\right|_D .
\end{equation}
 Differentiating this relation with respect to $t$ leads to
\begin{equation}
  \dot{m^2}_i= D_2 \gamma_i(a),
\end{equation}
where $D_2$ stands for a second order differential operator
(\ref{s}) introduced in Ref.~\cite{AKK}.

As  mentioned above one can make the same expansion not only in
equations,  but also in  solutions. Let us start with the simplest
case of pure gauge theory with one gauge coupling. Then one has in
a rigid theory
\begin{equation}
\int^{\displaystyle \alpha} \frac{d\alpha'}{\beta(\alpha')} =
 \log\left(\frac{Q^2}{\Lambda^2}\right).
\end{equation}
Making a substitution $\alpha \to \tilde{\alpha}$  and
 $\tilde{\Lambda}=\Lambda(1+c\theta^2+...)$ one has
\begin{equation}
\int^{\displaystyle  \tilde{\alpha}}
\frac{d\alpha'}{\beta(\alpha')}
 = \log\left(\frac{Q^2}{\tilde{\Lambda}^2}\right).
\end{equation}
 Expansion over $\eta$ gives
\begin{equation}
M= c \gamma(\alpha), \ \ \ \
\gamma(\alpha)=\frac{\beta(\alpha)}{\alpha}.
\end{equation}

One can make  the same  expansion for any analytic solution in a
rigid theory. Below we consider three particular examples, namely
the MSSM, the finite SUSY GUT and the Seiberg-Witten N=2 SUSY
model.

\section{Examples}

{\bf The MSSM}\vspace{0.3cm}

Consider the MSSM in low $\tan\beta$ regime.  One has three gauge
and one Yukawa
 coupling. The one-loop RG equations are~\cite{Ibanez}
\begin{eqnarray}
\dot{\alpha}_i&=&-b_i\alpha^2_i, \ \ \ \  b_i=(\frac{33}{5},1,-3),
\ \ i=1,2,3, \\
\dot{Y}_t&=&Y_t(\frac{16}{3}\alpha_3+3\alpha_2+\frac{13}{15}\alpha_1-6Y_t),
\end{eqnarray}
with the initial conditions: $\alpha_i(0)=\alpha_0, \ Y_t(0)=Y_0$
and $t=\ln(M_X^2/Q^2)$.
 Their solutions are given by~\cite{Ibanez}
\begin{equation}
\alpha_i(t)=\frac{\alpha_0}{1+b_i\alpha_0t}, \ \ \
Y_t(t)=\frac{Y_0E(t)}{1+6Y_0F(t)},
 \label{sol}
\end{equation}
where
\begin{eqnarray*}
E(t)&=&\prod_i(1+b_i\alpha_0t)^{\displaystyle c_i/b_i} , \ \ \
 c_i=(\frac{13}{15},3,\frac{16}{3}), \\
F(t)&=&\int^t_0 E(t')dt'.
\end{eqnarray*}

To get the solutions for the soft terms it is enough to perform
substitution $\alpha \to \tilde{\alpha}$ and $Y\to \tilde{Y}$ and
expand over $\eta $ and $\bar \eta$. Expanding the gauge coupling
in (\ref{sol}) up to $\eta$ one has (hereafter we assume
$M_{i0}=M_0$)
$$\alpha_iM_i=\frac{\alpha_0M_0}{1+b_i\alpha_0t}-\frac{\alpha_0
b_i\alpha_0M_0t}{(1+b_i\alpha_0t)^2}=\frac{\alpha_0}{1+b_i\alpha_0t}
\frac{M_0}{1+b_i\alpha_0t},$$ or
\begin{equation}
M_i(t)=\frac{M_0}{1+b_i\alpha_0t}.
\end{equation}
Performing the same expansion for the Yukawa coupling and using
the relations $$\left.
\frac{d\tilde{E}}{d\eta}\right|_\eta=M_0t\frac{dE}{dt}, \ \ \
\left.\frac{d\tilde{F}}{d\eta}\right|_\eta=M_0(tE-F),$$ one finds
a well known expression~\cite{Ibanez}
\begin{equation}
A_t(t)=\frac{A_0}{1+6Y_0F}+M_0\left( \frac{t}{E}\frac{dE}{dt}-
\frac{6Y_0}{1+6Y_0F}(tE-F)  \right). \label{a}
\end{equation}
To get the solution for the $\Sigma$ term one has to make
expansion over $\eta$ and $\bar \eta$. This can be done with the
help of the following relations
 $$\left.\frac{d^2\tilde{E}}{d\eta
d\bar \eta}\right|_{\eta,\bar \eta}=
M_0^2\frac{d}{dt}\left(t^2\frac{dE}{dt}\right), \ \ \
\left.\frac{d^2\tilde{F}}{d\eta d\bar \eta}\right|_{\eta,\bar
\eta} =M_0^2t^2\frac{dE}{dt}.$$ This leads to
\begin{equation}
\Sigma_t(t)=\frac{\Sigma_0-A_0^2}{1+6Y_0F}+\frac{(A_0-M_06Y_0(tE-F))^2}{(1+6Y_0F)^2}
+M_0^2\left[\frac{d}{dt}\left(\frac{t^2}{E}\frac{dE}{dt}\right)
-\frac{6Y_0}{1+6Y_0F}t^2\frac{dE}{dt}\right], \label{si}
\end{equation}
which is much simplier than known in the literature~\cite{Ibanez},
though coinciding with it after some cumbersome algebra.

With  analytic solutions (\ref{a},\ref{si}) one can analyze
asymptotics and, in particular, find the infrared quasi fixed
points~\cite{Hill} which correspond to $Y_0 \to \infty$
\begin{eqnarray}
Y^{FP}&=&\frac{E}{6F},  \label{Yf}\\
A^{FP}&=&M_0\left(\frac{t}{E}\frac{dE}{dt}-\frac{tE-F}{F}\right),
\label{Af}\\
\Sigma^{FP}&=&M_0^2\left[\left(\frac{tE-F}{F}\right)^2+\frac{d}{dt}
\left(\frac{t^2}{E}\frac{dE}{dt}\right)-\frac{t^2}{F}\frac{dE}{dt}\right].
\label{Sf}
\end{eqnarray}
However, the advantage of the Grassmannian expansion procedure is
that one can perform it for  fixed points as well. Thus the FP
solutions (\ref{Af},\ref{Sf}) can be directly obtained from a
fixed point for the rigid Yukawa coupling (\ref{Yf}) by
Grassmannian expansion. This explains, in particular, why fixed
point solutions for the soft couplings  exist if they exist for
the rigid ones and with the same stability properties~\cite{JJP}.

\vspace{0.6cm}

\noindent {\bf SUSY GUTs}\vspace{0.3cm}

One can consider not only  fixed points, but also more complicated
configurations like renormalization invariant trajectories which
lead to reduction of the couplings~\cite{JJ} or fixed lines or
surfaces~\cite{Schrempp},  or finiteness relations~\cite{Finite}.
The same procedure is valid here as well.

Let us consider, for example,  construction of a finite theory
(free from ultraviolet divergences) in the framework of SUSY GUTs.
It is achieved  in a rigid theory by a proper choice of the field
content and of the Yukawa couplings being the functions
 of the gauge one~\cite{Finite}
\begin{equation}
Y_k=Y_k(\alpha)= c_0^{(k)}\alpha+c_1^{(k)}\alpha^2 + . . . ,
\label{fin}
\end{equation}
where  coefficients $c_i^{(k)}$ are calculated within perturbation
theory.

To achieve complete finiteness, including the soft terms, one has
to choose the latter in a proper way~\cite{K}. To find it one just
have to  modify the  finiteness relation for the Yukawa coupling
(\ref{fin}) as
\begin{equation}
\tilde{Y}_k=Y_k(\tilde{\alpha}),
\end{equation}
and expand over $\eta$ and $\bar\eta$. This gives:
\begin{equation}
A_k=M\frac{d\ln Y_k}{d\ln\alpha},
\end{equation}
and  after the rearrangement of terms
\begin{equation}
\Sigma_k=M^2\frac{d}{d\alpha}\alpha^2\frac{d\ln Y_k}{d\alpha},
\end{equation}
which coincides with the relations found in Ref.~\cite{K}.

\vspace{0.6cm}

\noindent {\bf N=2 SUSY} \vspace{0.3cm}

Consider now the N=2 supersymmetric gauge theory. The Lagrangian
written in terms of N=2 superfields is~\cite{AG}:
\begin{equation}
{\cal L}=\frac{1}{4\pi}{\cal I}m Tr \int d^2\theta
d^2\tilde{\theta}
 \frac 12 \tau \Psi^2,
\end{equation}
where N=2 chiral superfield $\Psi(y,\theta,\tilde{\theta})$ is
defined by
 constraints $\bar{D}_{\dot{\alpha}}\Psi=0$ and $\bar{\tilde{D}}_{\dot{\alpha}}\Psi=0$
 and
\begin{equation}
\tau = i\frac{4\pi}{g^2}+\frac{\theta}{2\pi}^{\hspace{-0.1cm}
topological}\hspace{-1.1cm}.
\end{equation}

The expansion of $\Psi$  in terms of $\tilde{\theta}$  can be
written as
$$\Psi(y,\theta,\tilde{\theta})=\Psi^{(1)}(y,\theta)+\sqrt{2}\tilde{\theta}^\alpha
 \Psi^{(2)}_\alpha(y,\theta)+\tilde{\theta}^\alpha \tilde{\theta}_\alpha \Psi^{(3)}
 (y,\theta),$$
where $y^\mu=x^\mu+i\theta\sigma^\mu\bar \theta +i\tilde{\theta}
\sigma^\mu\bar{\tilde{\theta}}$  and   $\Psi^{(k)}(y,\theta)$ are
 N=1 chiral superfields.

The soft breaking of N=2 SUSY down to N=1 can be achieved by
shifting the imaginary part of $\tau$:
\begin{equation}
\tau  \to \tilde{\tau} =\tau +
i\frac{4\pi}{g^2}\tilde{\theta}^2M. \label{shift}
\end{equation}
This leads to
\begin{equation}
\Delta{\cal L}=\frac{1}{g^2} Tr \int d^2\theta \frac{M}{2}
(\Psi^{(1)})^2 ,
\end{equation}
which is the usual mass term for N=1 chiral superfield
$\Psi^{(1)}$ normalized to $1/g^2$.

Now one can use the power of duality in N=2 SUSY theory and take
the
 Seiberg-Witten solution~\cite{SW}
\begin{equation}
\tau = \frac{da_D}{du}/\frac{da}{du}, \label{SW}
\end{equation}
where
\begin{eqnarray*}
a_D(u)&=&\frac{i}{2}(u-1)F(1/2,1/2,2;\frac{1-u}{2}) ,\\
a(u)&=&\sqrt{2(1+u)}F(-1/2,1/2,1;\frac{2}{1+u}).
\end{eqnarray*}

Assuming that renormalizations in N=2 SUSY theory  follow the
properties of those in N=1 one can apply the same expansion
procedure. Substituting eq.(\ref{shift}) into (\ref{SW}) with $u
\to \tilde{u} =u(1+M_0\tilde{\theta}^2)$ and expanding over
$\tilde{\theta}^2$,
 one gets an analog of S-W solution for the mass term:
\begin{equation}
M=M_0\frac{\displaystyle {\cal I}m \left[u\left(\frac{a_D''}{a_D'}
-\frac{a''}{a'}\right)\tau \right]}{\displaystyle  {\cal I}m \
\tau}\, . \label{M}
\end{equation}
In perturbative regime ($u \sim Q^2/\Lambda^2 \to \infty$) one
has~\cite{AG} $a=\sqrt{2u}, \ a_D=\frac{\displaystyle
i}{\displaystyle \pi}a(2\ln a +1)$, which leads to
\begin{eqnarray*}
\frac{4\pi}{g^2}&=&\frac{1}{\pi}\left[\ln
Q^2/\Lambda^2+3\right],\\ M&=&M_0/\left[\ln
Q^2/\Lambda^2+3\right].
\end{eqnarray*}

This procedure can be continued introducing soft N=1 SUSY breaking
via $\theta$ dependent $\tau$ superfield. Thus one can achieve
soft SUSY breaking along the chain $$N=2 \ \  \Rightarrow \ \
N=1\ \ \Rightarrow \ \ N=0$$ preserving the  properties of the
exact solution. This will lead to a sequence of new solutions for
the soft terms like eq.(\ref{M}).

\section{Conclusion}

We conclude that the Grassmannian expansion in softly broken SUSY
theories happens to be a very efficient and powerful  method which
can be applied in various cases where the renormalization
procedure in concerned. It demonstrates once more that softly
broken SUSY theories  are contained in rigid  ones and inherit
their renormalization properties.

\vspace{1cm}

{\large \bf Acknowledgments}

\vspace{0.3cm}

Financial support from RFBR grants \# 96-02-17379a and \#
96-15-96030  and DFG grant \# 436 RUS  113/335 is kindly
acknowledged.


\end{document}